%% file: etop.tex
\documentstyle[preprint,aps,floats,psfig]{revtex}
\input{acp.com}
\begin{document}
\input{etop_0.tex}
\input{etop_1.tex}
\newpage
\bibliographystyle{ieeetr}
\bibliography{$HOME/porbib/porous,$HOME/tex/bib/publ}
\end{document}

%% file: etop_0.tex
\draft
\setcounter{page}{0}
\title{Local Percolation Probabilities
for a Natural Sandstone}
\author{R. Hilfe$\mbox{\rm r}^{1,2}$, 
T. Rag$\mbox{\rm e}^{3}$ and 
B. Virgi$\mbox{\rm n}^{3}$}
\address{
$\mbox{ }^1$ICA-1, Universit{\"a}t Stuttgart,
Pfaffenwaldring 27, 70569 Stuttgart\\
$\mbox{ }^2$Institut f{\"u}r Physik,
Universit{\"a}t Mainz,
55099 Mainz, Germany\\
$\mbox{ }^3$Institute of Physics,
University of Oslo,
0316 Oslo,
Norway}
\maketitle
\thispagestyle{empty}
\begin{abstract}
Local percolation probabilities are used to characterize the 
connectivity in porous and heterogeneous media.
Together with local porosity distributions they allow
to predict transport properties \cite{hil91d}.
While local porosity distributions are readily obtained, 
measurements of the local percolation probabilities are
more difficult and have not been attempted previously.
First measurements of three dimensional local porosity
distributions and percolation probabilities from a pore space 
reconstruction for a natural sandstone show that theoretical 
expectations and experimental results are consistent.
\end{abstract}

%% file: etop_1.tex
\newpage
According to the mixing law of local porosity theory 
\cite{hil91d,hil92a,hil92b,hil92f,hil93b,hil93c,hil94b,hil94g,hil95g,hil95d}
the effective frequency dependent dielectric function $\eps_e(\omega)$ of 
a heterogeneous mixture may be calculated by solving the integral equation
\begin{equation}
\int_0^1\!\!\frac{\eps_C(\omega;\phi)-\eps_e(\omega)}
{\eps_C(\omega;\phi)+2\eps_e(\omega)}\lambda(\phi)\mu(\phi)d\phi +
\int_0^1\!\!\frac{\eps_B(\omega;\phi)-\eps_e(\omega)}
{\eps_B(\omega;\phi)+2\eps_e(\omega)}(1-\lambda(\phi))\mu(\phi)d\phi = 0
\label{LPT1}
\end{equation}
where 
\begin{eqnarray}
\eps_C(\omega;\phi) & = & \eps_1(\omega)\left( 1 -
\frac{1-\phi}{(1-\eps_2(\omega)/\eps_1(\omega))^{-1}-\phi /3} \right) 
\label{epsc} \\
\eps_B(\omega;\phi) & = & \eps_2(\omega)\left( 1 -
\frac{\phi}{(1-\eps_1(\omega)/\eps_2(\omega))^{-1}-(1-\phi)/3} \right)
\label{epsb}
\end{eqnarray}
In (\ref{LPT1}) the local porosity distribution $\mu(\phi)$ and the local 
percolation probability $\lambda(\phi)$ are geometrical input functions
which will be defined shortly.
The functions $\eps_1(\omega)$, and $\eps_2(\omega)$
are the frequency dependent dielectric functions of the pure materials.
If $\eps_i(\omega), i=1,2$ and $\mu(\phi),\lambda(\phi)$ are known
then (\ref{LPT1}) gives a prediction for the 
effective dielectric function $\eps_e(\omega)$.

Measurements of the local porosity distribution $\mu(\phi)$ are readily
obtained from thin sections \cite{hil93c,hil94g,NHH92}.
Measurements of the local percolation probabilities $\lambda(\phi)$,
on the other hand, are more difficult, and have not been reported.
The main purpose in this paper is to present preliminary measurements
of local percolation probabilities for a natural sandstone specimen.

Define $\PP$ to be the pore space of a two component porous medium.
The complement of $\PP$ in $\RR^3$ is called the matrix space $\MM$.
More generally $\PP$ and $\MM$ may represent the two components in
a heterogeneous medium.
The local porosity $\phi(\KK)$ measured within a measuring region $\KK$ 
is defined by
\begin{equation}
\phi(\KK)=\frac{V(\PP\cap\KK)}{V(\KK)}
\end{equation}
where $V(\KK)$
is the volume of a set $\KK$.
Let $\ggg_i\in\ZZ^3$ denote the lattice vectors of a simple cubic 
lattice, and let
\begin{equation}
\KK(\rr,L)=\{\rr^\prime\in\RR^3:|r^\prime_i-r_i|\leq L/2, i=1,2,3\}
\end{equation}
denote a cube of sidelength $L$ centered at $\rr$.

Given the notation introduced above the local porosity distribution
may be defined as \cite{hil95d}
\begin{equation}
\mu(\phi,L)=\lim_{M\rai}\frac{1}{M}
\sum_{i=1}^M\delta(\phi-\phi(\KK(L\ggg_i,L)))
\label{empiriclpd}
\end{equation}
if the limit exists. Here $\delta(x)$ is the Dirac $\delta$-function.
In practice the sample is finite, and hence the limiting process
terminates after a finite number $M$ of measurement cells.
The resulting histogram is used as an approximation for $\mu$.

Local porosity distributions quantify the fluctuations in volume
fractions.
To characterize the transport properties,however, it is necessary 
to quantify the degree of connectedness of the porous
medium. Two points inside the pore space $\PP$ are called
connected if there exists a path entirely within the
pore space that connects the two points.
A cubic measurement cell $\KK_j$ in a cubic partitioning
is called percolating in the $x$-direction if there exists
a path within $\PP\cap\KK_j$ connecting those two opposite
faces of $\KK_j$ that are perpendicular to the $x$-direction.
Percolation in the $y$- or $z$-direction is defined analogously.
The local percolation probability $\lambda_x(\phi,L)$
in the $x$-direction is defined as the fraction of 
measurement cells that are percolating in the
$x$-direction and have a local porosity $\phi$.
The local percolation probabilities $\lambda_y(\phi,L)$
and $\lambda_z(\phi,L)$ are defined analogously.

\begin{figure}
\centerline{
\psfig{file=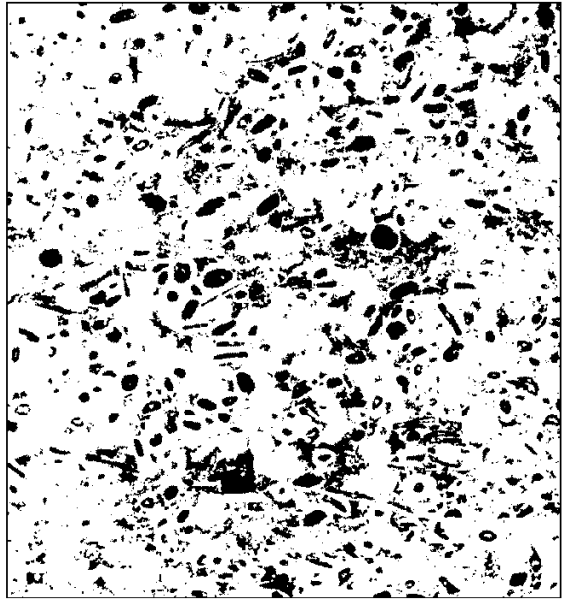,height=12.5cm}
}
\vspace*{12pt}
\caption{
Digitized thin section image of Savonnier oolithic sandstone
\protect\cite{OVHKM95}.
The pore space $\PP$ is coloured black, the matrix space is 
rendered white.}
\label{grid}
\end{figure}

To measure $\mu(\phi,l)$ and $\lambda(\phi,L)$ in practice
it is necessary to reconstruct the three dimensional pore
space $\PP$ for digital processing.
This was done by the method of serial sectioning for
a Savonnier oolithic sandstone specimen \cite{ost95,OVHKM95}.
This type of sandstone exhibits oomoldic porosity with
ellipsoidal pores and grains of sizes between $200\mu$m and $300\mu$m
\cite{ost95}.
After filling the stone with a coloured epoxy resin
the specimen was cut and polished.
The polished surface was photographed before removing
another layer of material parallel to first surface.
The second surface was then polished and photographed
before iterating the procedure.
A total of 99 sections was prepared.
The distance between consecutive planes was about $10\mu$m.
Each photograph represented an area of roughly $2$cm$\times 2$cm.
The photographs were scanned and thresholded into a binary image
with $1904\times 1904$ pixels.
This corresponds to a resolution within each image of roughly $10\mu$m.
Figure \ref{grid} shows a typical binary image from within one of
the 100 planes measuring roughly $1$cm along each side.

\begin{figure}
\centerline{
\psfig{file=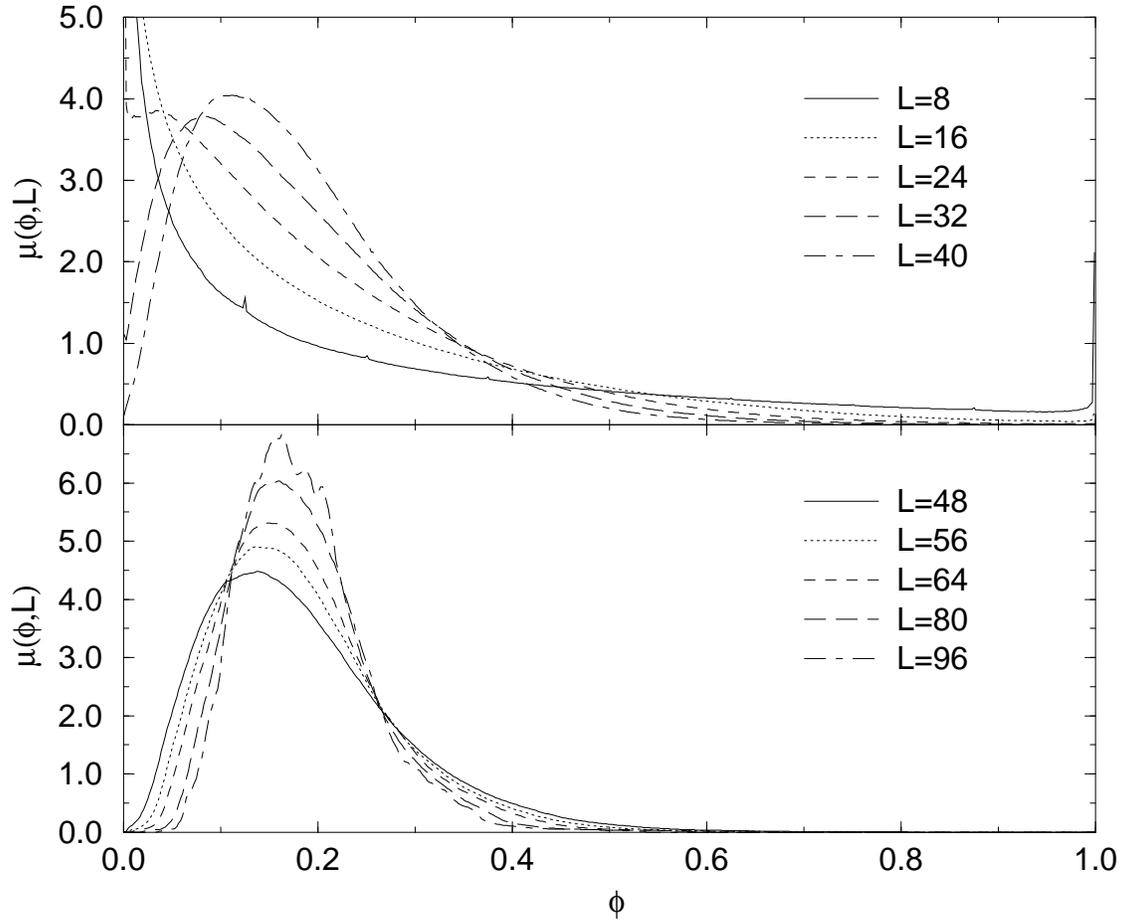,height=12.5cm}
}
\vspace*{12pt}
\caption{Local porosity distributions $\mu(\phi;L)$ for a cubic
measurement cells of sidelengths $L$. 
The values of $L$ for the different curves are indicated in the legend.}
\label{lpds}
\end{figure}

The family of three dimensional local porosity distributions 
$\mu(\phi,L)$ for this sample is obtained straightforwardly. 
The results are displayed in Figure \ref{lpds} for various $L$. 
Ref. \cite{hil95g} has investigated the question to what extent 
the three dimensional local porosity distribution can be measured
from two dimensional sections. 
While this is not generally possible, it was found that a two
dimensional measurement combined with an appropriate rescaling 
of the length $L$ gives good results.

\begin{figure}
\centerline{
\psfig{file=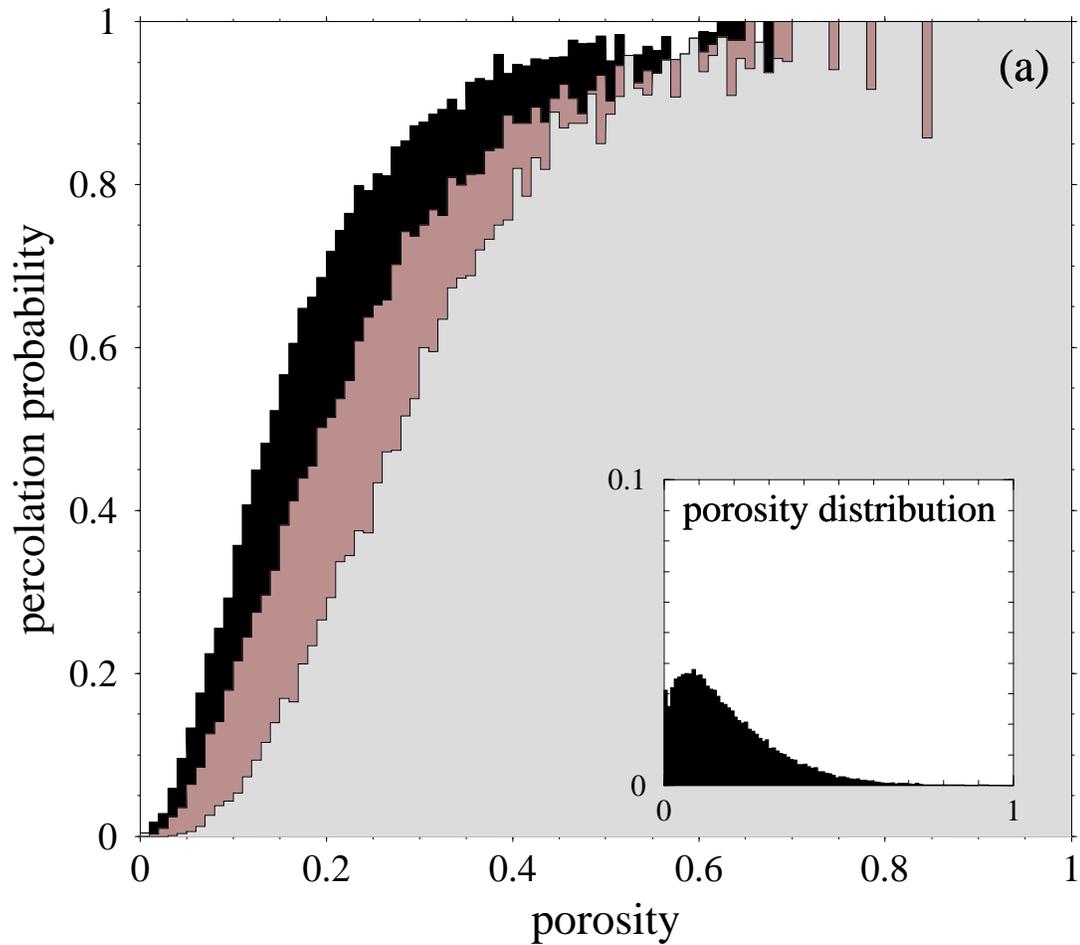,height=12.5cm}
}
\caption{
Local percolation probability functions $\lambda_x(\phi,L)$ (black),
$\lambda_y(\phi,L)$ (medium gray), and $\lambda_z(\phi,L)$ (light gray)
for the Savonnier oolithic sandstone shown in Figure \protect\ref{grid} 
using a simple cubic lattice of measurement cells with side lengths
(lattice constant) $L=32$.
The local porosity distribution $\mu(\phi,L)$ for $L=32$ is shown as 
the inset.}
\label{lpps}
\end{figure}

A preliminary measurement of the local percolation
probabilities has also been carried out for the oolithic
sandstone specimen.
The results are shown in Figure \ref{lpps} for cubic
measurement cells of size $L=32$.
The curve $\lambda_x(\phi;32)$ is shown in black,
the curve $\lambda_y(\phi;32)$ is shown in medium gray,
and the curve $\lambda_z(\phi;32)$ is shown in light gray.
The local porosity distribution for $L=32$ is reproduced
as the inset.

While the data show much scatter these preliminary estimates 
for the local percolation probabilities indicate that the 
sandstone has anisotropic connectivity.
The same trend is observed for smaller as well as for
larger values of $L$, although the strength of the
anisotropy appears to change slightly.

The data are in agreement with the theoretical expectation
that the local percolation probabilities should increase
monotonically from $0$ to $1$.
The shape of the curves conforms to the one expected for
systems with a percolation threshold.
An example is the grain consolidation model with random 
packings \cite{hil91d}.
For small values of $L$, however, the shape of the local
percolation probabilities resembles that for the central 
pore model \cite{hil91d}.
All of these conclusions are very preliminary.
More work is necessary to corroborate them, and to 
validate the measurement.

ACKNOWLEDGEMENT:
The authors are grateful to Chr. Ostertag-Henning,
E. Haslund, Dr. U. Mann, Prof.Dr. B. N{\o}st, Prof.Dr.
D.H. Welte and Prof.Dr. R. Koch for discussions, and
to Norges Forskningsr{\aa}d and the Forschungszentrum 
J{\"u}lich for partial financial support.

\newpage
\vspace*{2cm}
\newpage
\vspace*{2cm}
\newpage
\vspace*{2cm}
